\documentclass[12pt]{iopart}
\usepackage{graphics}

\begin{document}

\title[On the Cyclotomic Quantum Algebra of  Time Perception]
{On the Cyclotomic Quantum Algebra\\ of Time Perception}

\author{Michel Planat\dag\ \footnote[3]{To whom correspondence should be addressed
Tel: (33) 3 81 85 39 57; Fax: (33) 3 81 85 39 98; planat@lpmo.edu}
}

\address{\dag\ Institut FEMTO-ST, D\'{e}partement LPMO,\\ 32 Avenue de l'Observatoire, 25044 Besan\c{c}on Cedex,
France\\}

%\maketitle

\begin{abstract}

I develop the idea that time perception is the quantum counterpart
to time measurement. Phase-locking and prime number theory were
proposed as the unifying concepts for understanding the optimal
synchronization of clocks and their $1/f$ frequency noise. Time
perception is shown to depend on the thermodynamics of a quantum
algebra of number and phase operators already proposed for quantum
computational tasks, and to evolve according to a Hamiltonian
mimicking Fechner's law. The mathematics is Bost and Connes
quantum model for prime numbers. The picture that emerges is a
unique perception state above a critical temperature and plenty of
them allowed below, which are parametrized by the symmetry group
for the primitive roots of unity. Squeezing of phase fluctuations
close to the phase transition temperature may play a role in
memory encoding and conscious activity.

\end{abstract}

\maketitle
%\pacs{ 03.67.-a, 05.40.Ca, 02.10.De, 02.30.Nw}


\section{Introduction}

In a recent issue of this journal B. Flanagan \cite{Flanagan03}
discusses parallels between the perception of color and its
concomitant quantum field. We follow that line of reasoning by
proposing the concept of phase-locking to parallel classical
measurements on oscillators with the quantum perception of time.
Period measurements of a test oscillator against a reference one
is usually performed close to baseband, thanks to a non-linear
mixing element and a low pass filter. The resulting beat note in
units of the frequency of the reference oscillator results from
the continued fraction expansion of the input frequency ratio. The
beat frequency exhibits variability, with $1/f$ power spectrum,
that we explained from phase-locking of the input oscillators. We
could model the effect by considering a discrete coupling
coefficient versus time, related to the logarithm of prime numbers
and also to the Riemann zeta function and its critical zeros
\cite{Planat03}.

Time evolution in human classical oscillators such as the
circadian rhythm in plants, the heart rate of melatonine secretion
should obey the same rules because they are slaved to the
lightning environment or to internal pacemakers. But does time
perception resort to the arithmetic above? Our postulate is that
our mind still uses phase-locking, but in a discrete algebraic
way, from quantum finite fields lying in the brain. The object
under control by mental states would be
$\mathcal{Z}_q=\mathcal{Z}/q\mathcal{Z}$, the ring of integers of
modulo $q$, with $q$ the finite dimension of the quantum field. In
particular we claim the ability of our mind to lock to the largest
cyclic subgroup in $\mathcal{Z}_q$. The human ability to perceive
the greatest common divisor in the frequencies of two sounds,
instead of their beat frequency, is well known, as is the ability
to implicitly manage with continued fraction expansions in the
musical design of well tempered scales \cite{Schroeder99}. On
classical computers these tasks require a polynomial time. In
contrast, finding the primitive roots of an algebraic equation
$a^{\alpha}=1$($\rm{mod}$$~q)$, on which periodicity query
depends, requires exponential time. Thus our intuitive sense of
time, of prime numbers, of primitive roots should result from the
ability of our mind to perform some sort of quantum computation.
So the $1/f$ noise effects observed in human cognition
\cite{Gilden01} would be related to the $1/f^{\gamma}$ noise
observed in $Z_q$ about the period of its largest cyclic subgroup.

 The best possible introduction to our research
is still in the visionary Poincar\'{e} words \cite{Poincare05}:

{\it The Physical Continuum. —We are next led to ask if the idea
of the mathematical continuum is not simply drawn from experiment.
If that be so, the rough data of experiment, which are our
sensations, could be measured. We might, indeed, be tempted to
believe that this is so, for in recent times there has been an
attempt to measure them, and a law has even been formulated, known
as Fechner's law, according to which sensation is proportional to
the logarithm of the stimulus. But if we examine the experiments
by which the endeavor has been made to establish this law, we
shall be led to a diametrically opposite conclusion. It has, for
instance, been observed that a weight $A$ of $10$ grammes and a
weight $B$ of $11$ grammes produced identical sensations, that the
weight $B$ could no longer be distinguished from a weight C of 12
grammes, but that the weight $A$ was readily distinguished from
the weight $C$. Thus the rough results of the experiments may be
expressed by the following relations: $A=B, B=C, A<C$, which may
be regarded as the formula of the physical continuum. But here is
an intolerable disagreement with the law of contradiction, and the
necessity of banishing this disagreement has compelled us to
invent the mathematical continuum. We are therefore forced to
conclude that this notion has been created entirely by the mind,
but it is experiment that has provided the opportunity. We cannot
believe that two quantities which are equal to a third are not
equal to one another, and we are thus led to suppose that $A$ is
different from $B$, and $B$ from $C$, and that if we have not been
aware of this, it is due to the imperfections of our senses.

 The Creation of the Mathematical Continuum: First Stage. — So far it would suffice, in
order to account for facts, to intercalate between $A$ and $B$ a
small number of terms which would remain discrete. What happens
now if we have recourse to some instrument to make up for the
weakness of our senses? If, for example, we use a microscope? Such
terms as $A$ and $B$, which before were indistinguishable from one
another, appear now to be distinct: but between $A$ and $B$, which
are distinct; is intercalated another new term $D$, which we can
distinguish neither from $A$ nor from $B$. Although we may use the
most delicate methods, the rough results of our experiments will
always present the characters of the physical continuum with the
contradiction which is inherent in it. We only escape from it by
incessantly intercalating new terms between the terms already
distinguished, and this operation must be pursued indefinitely. We
might conceive that it would be possible to stop if we could
imagine an instrument powerful enough to decompose the physical
continuum into discrete elements, just as the telescope resolves
the Milky Way into stars. But this we cannot imagine; it is always
with our senses that we use our instruments; it is with the eye
that we observe the image magnified by the microscope, and this
image must therefore always retain the characters of visual
sensation, and therefore those of the physical continuum.}

The goal of the paper is to connect Poincar\'{e} physical
continuum to the remarkable quantum phase model due to J.B. Bost
and A. Connes \cite{Bost95}: {\it In this paper we construct a
$C^*$-dynamical system whose partition function is the Riemann
$\zeta$ function. It admits the $\zeta$ function as partition
function and the Galois group
$Gal(\mathcal{Q}^{\rm{cycl}}/\mathcal{Q})$ of the cyclotomic
extension $\mathcal{Q}^{\rm{cycl}}$ of $\mathcal{Q}$ as symmetry
group. Moreover, it exhibits a phase transition with spontaneous
symmetry breaking at inverse temperature $\beta=1$. The original
motivation for these results comes from the work of B. Julia.}

\section{Classical phase-locking}
\label{classical}

Poincar\'{e} physical continuum can be figured out by the concept
of phase-locking of clocks. The oscillators in our body, from the
skin to the nerve cells, are engaged in our sensation of weight.
Two groups of oscillators compare the physical strengths of the
stimuli and react as $A$ and $B$ by locking at values
$\frac{A}{B}=\frac{1}{1}$, or $\frac{A}{B}=\frac{1}{2}\cdots$, or
more generally at some simple rational ratio
$\frac{A}{B}=\frac{p}{q}$, where $p$ and $q$ are coprime to each
other, i.e. with greatest common divisor $(p,q)=1$. One model of
the phase-locking phenomenon between coupled oscillators of
angular frequency $\omega_0$ and $\omega$ is the Adler's equation
\cite{Planat03}
\begin{equation}
\frac{d\Phi}{dt}+K \sin \Phi= \Delta \omega. \label{Adler}
\end{equation}
It is achieved for example from a non-linear phase detector using
electronic diode mixers and a sustaining loop which shifts the
frequency of a voltage controlled oscillator. The same physical
mechanism also occurs in phase-locked lasers and in feedbacks with
biologic oscillators. The symbol $\Phi$ in (\ref{Adler})  is the
phase difference between the oscillators, $\Delta
\omega=\omega-\omega_0$ is the detuning frequency and $K$ is the
strength of the interaction. This model leads to a mean frequency
$\langle  \dot{\Phi} \rangle=0$ which is zero inside the
phase-locked zone $|\Delta \omega|<2K$ and which is $\langle
\dot{\Phi} \rangle=\Delta \omega(1-K^2/(\Delta \omega)^2)^{1/2}$
outside that zone. Note that if the oscillators have frequencies
far apart, then $\langle \dot{\Phi} \rangle \rightarrow \Delta
\omega$, so that the strength of \lq\lq sensation" is proportional
to the \lq\lq physical" strength. At that stage the model can only
discriminate if two physical strengths are equal or not. For a
better operation of phase-locking the harmonics generated in the
interaction have to be accounted for. Experimentally one finds
$1/f$ frequency fluctuations about the unperturbed signal. Their
range can be found easily by differentiating the shifted frequency
\cite{Planat03}. Technical details of the frequency lockings
between interacting harmonics need to have recourse to the theory
of continued fraction expansions. A simple model of the
phase-lockings is the so-called Arnold map
\begin{equation}
\Phi_{n+1}=\Phi_n+2\pi\Omega-c~ \sin \Phi_n, \label{Arnold}
\end{equation}
where the coupling coefficient is $c=\frac{K}{\omega_0}$. Such a
non-linear map is studied by introducing the winding number
$\nu=\lim _{n \rightarrow \infty } (\Phi_n-\Phi_0)/(2 \pi n)$. The
limit exists everywhere as long as $c<1$. The curve $\nu$ versus
$\Omega$ is a devil's staircase with steps attached to the
rational values $\Omega=\frac{p}{q}$ and the width increases with
the coupling coefficient $c$. The structure in steps corresponds
to the Poincar\'{e} physical continuum.

It is well known that the phase-locking zones may overlap if $c>1$
leading to chaos from quasi-periodicity \cite{Planat03}. However,
this type of chaos doesn't resort to $1/f$ noise since the
phase-locked loops generally operate at quite small values of the
coupling $c$. To account for the $1/f$ noise we refined Arnold's
equation by introducing a discrete time dependance on $c$ able to
reflect the tiny energy exchanges occurring from time to time and
needed for phase synchronization. We postulated a modulation of
$c$ by the Mangoldt function $\Lambda(n)$ of prime number theory.
It is defined as
\begin{equation}
\Lambda(n)=\left\{\begin{array}{ll}
\ln b &~~\mbox{if}~ n=b^k,~b~\mbox{a~prime},\\
 0 &~~ \mbox{otherwise}.\\
\end{array}\right.
\label{Mangoldt}
\end{equation}
Mangoldt function arises in prime number theory as the logarithmic
derivative of Riemann zeta function $\zeta(s)=\sum_{n\ge
1}n^{-s}$, $\Re (s)\ge1$ since $d \ln \zeta(s)/ds=\sum_{n\ge
1}\Lambda(n) n^{-s}$. Riemann zeta function can be extended
analytically to the whole plane of the complex variable $s$,
except for a pole at $s=1$. The zeta function also shows trivial
zeros on the negative axis, at $s=-2l$, $l$ integer, and an
infinite number of very irregularly spaced zeros on the vertical
axis $\Re{s}=\frac{1}{2}$. Riemann hypothesis states that indeed
all zeros are located on that critical axis. In addition the
averaged Mangoldt function can be expressed explicitly in terms of
critical zeros. Numerically the averaged Mangoldt function
deviates from $1$ with a random looking term, having a $1/f$ type
power spectral density. This fact motivated us to put it as a
modulating term of $c$. In Sect.\ref{PhaseTransition} below, it
will be shown how the pole of the Riemann zeta function is
associated with the critical temperature $T_c$ in the phase
transition model of time perception, and how  phase fluctuations
of the Mangoldt function type occur close to $T_c$.

\section{Quantum phase-locking}
\subsection{Quantum phase states}

Our hypothesis is that mental states of human perception still
rely on phase-locking but with an important feature added: the
discreteness of phase states, which adds to the discreteness of
frequency ratios between the interacting oscillators. What
characterizes best the quantum time perception in contrast to
classical time measurement is the existence of quantum operators
sustaining this algebra of phase states, and a properly chosen
evolution Hamiltonian operator. Otherwise our approach is quantum
field theory and quantum statistical mechanics.

Let us start with the familiar Fock states of the quantized
electromagnetic field (the photon occupation states) $|n \rangle$
which live in an infinite dimensional Hilbert space. They are
orthogonal to each other: $\langle n|m\rangle
 =\delta_{mn}$, where $\delta_{mn}$ is the Dirac symbol. The states form a
 complete set: $\sum_{n=0}^{\infty}|n\rangle \langle n|=1$.

 The annihilation operator $\hat{a}$ removes one photon from the electromagnetic field
\begin{equation}
\hat{a}|n\rangle=\sqrt{n}|n-1\rangle,~ n=1,2,\cdots
\label{annihilation}
\end{equation}
Similarly the creation operator $\hat{a}^{\dag}$ adds one photon:
$\hat{a}^{\dag}|n\rangle=\sqrt{n+1}|n+1\rangle$, $n=0,1,\cdots$
These operators follow the commutation relation
$[\hat{a},\hat{a}^{\dag}]=1$. The operator $N=\hat{a}^{\dag}
\hat{a}$ represents the particle number operator and satisfies the
eigenvalue equation $N|n\rangle=n|n\rangle$.

The algebra can be simplified by removing the normalization factor
$\sqrt{n}$ in (\ref{annihilation}) defining thus the shift
operator $E$ as
\begin{equation}
E |n\rangle =|n-1\rangle,~ n=1,2,\cdots \label{phaseop}
\end{equation}
Similarly $E^{\dag}|n\rangle=|n+1\rangle$, $n=0,1,\cdots$. The
operator $E$ has been known as the (exponential) phase operator
\begin{equation}
E=e^{i\Phi}=(N+1)^{-1/2}\hat{a}=\sum_{n=0}^{\infty}|n\rangle\langle
n+1|. \label{expophase}
\end{equation}
The problem with this definition is that the operator $E$ is only
one-sided unitary since $E E^{\dag}=1$,
$E^{\dag}E=1-|0\rangle\langle0|$, due to the vacuum-state
projector $|0\rangle\langle0|$. The problem of defining a
Hermitian quantum phase operator was already encountered by the
founder of quantum field theory P.A.M. Dirac \cite{Lynch95}. The
eigenvectors of $E$ have been known as the Susskind-Glogower phase
states; they satisfy the eigenvalue equation
\begin{equation}
E|\psi\rangle=e^{i\psi}|\psi\rangle, ~~$\rm{with}$~
|\Psi\rangle=\sum_{n=0}^{\infty}e^{in\psi}|n\rangle.
\label{Susskind}
\end{equation}
They are non-orthogonal to each other and form an overcomplete
basis which solves the identity operator since
$\frac{1}{2\pi}\int_{-\pi}^{\pi}d\psi|\psi\rangle\langle \psi|=1$.
When properly normalized the phase states  $|\psi\rangle$ in
(\ref{Susskind}) are a special case of the $SU(1,1)$ Perelomov
coherent states \cite{Vourdas93},\cite{Perelomov86}. The operator
$\cos \Phi=\frac{1}{2}(E+E^{\dag})$ is used in the theory of
Cooper pair box with a very thin junction when the junction energy
$E_J \cos \Phi$ is higher than the electrostatic energy
\cite{Bouchiat98}.

Further progress in the definition of phase operator was obtained
by Pegg and Barnett \cite{Lynch95} in the context of a finite
Hilbert space. From now we consider a finite basis $|n\rangle$ of
number states, with the index $n \in \mathcal{Z}_q =\mathcal{Z}/q
\mathcal{Z}$, the ring of integers modulo $q$. The phase states
are now defined from the discrete Fourier transform (or more
precisely from the quantum Fourier transform since the
superposition is on Fock states not on real numbers)
\begin{equation}
|\theta_p\rangle=\frac{1}{\sqrt{q}}\sum_{n=0}^{q-1}\exp(2i\pi\frac{p}{q}n)|n\rangle.
\label{QFT}
\end{equation}
The states are eigenstates of the Hermitian phase operator
$\Theta_q$ such that
$\Theta_q|\theta_p\rangle=\theta_p|\theta_p\rangle$ where
$\Theta_q=\sum_{p=0}^{q-1}\theta_p |\theta_p\rangle\langle
\theta_p|$ and $\theta_p =\theta_0+2\pi p/q$ with $\theta_0$ a
reference angle. It is implicit in the definition (\ref{QFT}) that
the Hilbert space is of finite dimension $q$. The states
$|\theta_p\rangle$ form an orthonormal set and in addition the
projector over the subspace of phase states is
$\sum_{p=0}^{q-1}|\theta_p\rangle\langle \theta_p|=1_q$, where
$1_q$ is the unitary operator. Given a state $|F\rangle$ one can
write a probability distribution $|\langle \theta_p|F\rangle|^2$
which may be used to compute various moments, e.g. expectation
values, variances. The key element of the formalism is that first
the calculations are done in the subspace of dimension $q$, then
the limit $q \rightarrow \infty$ is taken.

\subsection{Primitive roots and $1/f$ noise}

Let us consider the ring $\mathcal{Z}_q$. An important point about
this set is that it is endowed with cyclic properties. If $q=p$, a
prime number, then the period is $p-1$; in any case they are
cyclic subgroups in $\mathcal{Z}_q$. This can be shown by the
examination of the algebraic equation
\begin{equation}
a^{\alpha} = 1($\rm{mod}$~q). \label{power}
\end{equation}
Let us ask the question: what is the cycle of largest period in
$\mathcal{Z}_q$? To find it one should look at the primitive roots
which are defined as the solution of (\ref{power}) such that the
equation is wrong for any $1 \le \alpha < q-1$ and true only for
$\alpha=q-1$. If $q=b$, a prime number, and $b=7$, the largest
period is thus $\phi(b)=b-1=6$, and the cycle is as given in Table
1.
\begin{table}[htbp]
\caption{$(Z/7Z)^*$ is a cyclic group of order $\phi(7)=6$.}
\begin{center} \footnotesize
\begin{tabular}{|c|c|c|c|c|c|c|c|c|}
\hline {$\alpha$}     & {1} &{2} &{3} &{4} &{5} &{6} &{7} &{8}  \\
\hline {$3^{\alpha}$} & {3} &{2} &{6} &{4} &{5} &{1} &{3} &{2}\\
\hline
\end{tabular}
\end{center}
\end{table}
 If $q=2$, or $4$, or $q=b^r$, a power of a prime number $>2$, or $q=2b^r$, twice the
power of a prime number $>2$, then a primitive root exists, and
the largest cycle in the group is $\phi(q)$. For example $a=2$ and
$q=3^2$ leads to the period $\phi(9)=6<q-1=8$, as it is shown in
Table 2.

\begin{table}[htbp]
\caption{$(Z/3^2 Z)^*$,  is a cyclic group of order $\phi(9)=6$.}
\begin{center}
\footnotesize
\begin{tabular}{|c|c|c|c|c|c|c|c|c|}
\hline {$\alpha$}     & {1} &{2} &{3} &{4} &{5} &{6} &{7} &{8}  \\
\hline {$2^{\alpha}$} & {2} &{4} &{8} &{7} &{5} &{1} &{2} &{4}\\
\hline
\end{tabular}
\end{center}
\end{table}

Otherwise there is no primitive root. The period of the largest
cycle in $\mathcal{Z}_q=\mathcal{Z}/q\mathcal{Z}$ can still be
calculated and is called the Carmichael Lambda function
$\lambda(q)$. It is shown in Table 3 for the case $a=3$ and $q=8$.
It is $\lambda(8)=2<\phi(8)=4<8-1=7$.

\begin{table}[htbp]
\caption{$(Z/8Z)^*$ has a largest cyclic group of order
$\lambda(8)=2$.}
\begin{center}
\footnotesize
\begin{tabular}{|c|c|c|c|c|c|c|c|c|}
\hline {$\alpha$}     & {1} &{2} &{3} &{4} &{5} &{6} &{7} &{8}  \\
\hline {$3^{\alpha}$} & {3} &{1} &{3} &{1} &{3} &{1} &{3} &{1}\\
\hline
\end{tabular}
\end{center}
\end{table}
\begin{figure}[ht]
\centering{\resizebox{8cm}{!}{\includegraphics{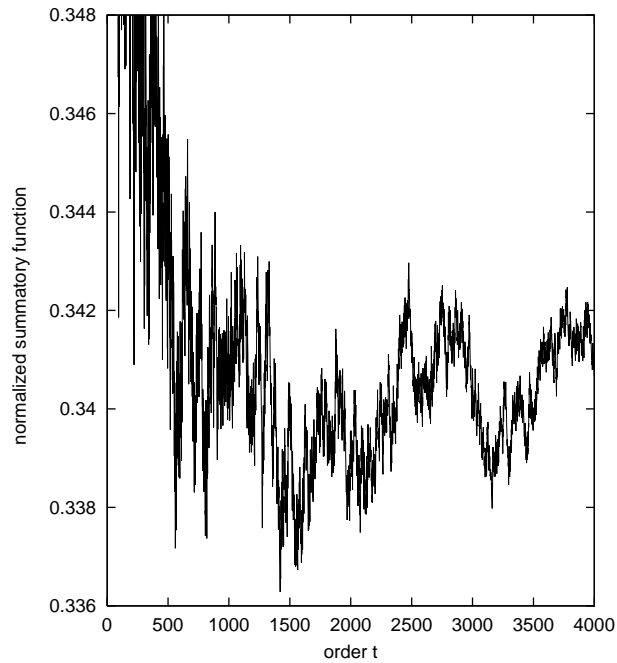}}}
\caption{Normalized Carmichael lambda function: $(\sum_1^t
\lambda(n))/t^{1.90}$.}
\end{figure}
\begin{figure}[ht]
\centering{\resizebox{8cm}{!}{\includegraphics{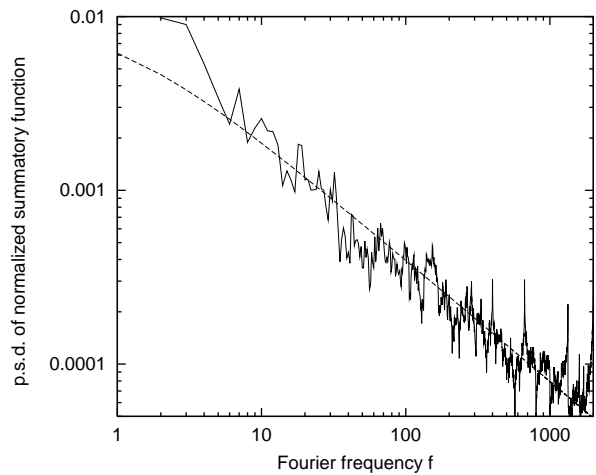}}}
\caption{FFT of the normalized Carmichael lambda function. The
straight line has slope $-0.70$.}
\end{figure}

Fig. 1 shows the properly normalized period for the cycles in
$\mathcal{Z}_q=\mathcal{Z}/q\mathcal{Z}$. Its fractal character
can be appreciated by looking at the corresponding power spectral
density (FFT) shown in Fig. 2. It has the form of a $1/f^{\gamma}$
noise, with $\gamma=0.70$. For a more refined link between
primitive roots, cyclotomy and Ramanujan sums see also
\cite{Moree03}.

Let us remind that the algebraic equation (\ref{power}) is used as
a numerical trap door for public key encryption (the RSA
cryptosystem) \cite{Schroeder99}, p.125. While it is easy to
multiply $1000$-digit numbers within a time of few milliseconds on
a modern classical computer, the opposite task of factorizing is
impossible in a reasonable time. This impossibility would be
lifted on a quantum computer thanks to Schor's algorithm of
finding periods efficiently in $\mathcal{Z}_q$ (see
Sect.\ref{PhaseTransition}).

\subsection{Quantum computation and the Bost and Connes algebra}
\label{PhaseTransition}

The quantum operator corresponding to (\ref{power}) is another
shift $\mu_a$ in the space of number states, where $a\in
\mathcal{Z}_q$ and $(a,n)=1$. It is defined as
\begin{equation}
\mu_a|n>=|an($\rm{mod}$~q)\rangle, \label{shift2}
\end{equation}
It is multiplicative: $\mu_k\mu_l=\mu_{kl}$ and $\mu_k^*\mu_k=1$,
$k,l\in \mathcal{N}$. The eigenvalues and eigenvectors of the
operator (\ref{shift2}) allow to define the order $r=ord_q(a)$ of
$a~($\rm{mod}$)~ q$ which is the smallest exponent such that
(\ref{power}) is  satisfied.  When the order is $r=\phi(q)$ then
$a$ is a primitive  root. This results from Euler theorem which
states that  $a^{\phi(q)}=1($\rm{mod}$~ q)$. If the order is the
same for any $a$ such that $(a,q)=1$ it is also called a universal
exponent. Thus $\phi(q)$ is a universal exponent. It is the
smallest one if $a$ is a primitive root, otherwise there is a
smaller one which is Carmichael lambda function introduced above.
Let us summarize
\begin{equation}
ord_q(a)\le\lambda(q)\le\phi(q)\le q-1. \label{inequality}
\end{equation}
The eigenvalues of operator (\ref{shift2}) are of the form
$\exp(2i \pi\frac{k}{r})$ and the corresponding eigenvectors are
given as a quantum Fourier transform \cite{Ekert00}, p. 22.
\begin{equation}
|u_k\rangle=
\frac{1}{\sqrt{r}}\sum_{j=0}^{r-1}\exp(-2i\pi\frac{kj}{r})|a^j$\rm{mod}$~
q \rangle. \label{QFT2}
\end{equation}
We observe that (\ref{QFT2}) acts on the subspace of dimension $r$
(the least period) from the powers of $a$ in (\ref{power}). One
sees that the $u_k$'s acquire the status of phase states in the
cyclic subspace of $Z_q$ of dimension r.

The operator $\mu_a$ is used in the context of quantum computation
in relation to Schor's algorithm \cite{Ekert00}. The goal is to
factor integers efficiently (in polynomial time instead of the
exponential time needed on the classical computer). It is in fact
related to the efficient estimation of the period $r$ taken from
the eigenvalue $\exp(2i \pi\frac{k}{r})$, and to the efficient
implementation of quantum Fourier transforms on the quantum
computer.

The shift operator (\ref{shift2}) which quantizes (\ref{power})
plays the role of multiplication in the language of operators. The
addition operator is easier to define
\begin{equation}
\e_p|n\rangle=\exp(2i\pi\frac{p}{q}n)|n\rangle. \label{phaseop2}
\end{equation}
It encodes the individuals in the quantum Fourier transform
(\ref{QFT}). These individuals are eigenvalues of the operator
$e_p$. One gets $e_0=1$, $e_p^*=e_{-p}$, $e_le_m=e_{l+m}$. Both
operators $\mu_a$ and $e_p$  form an algebra $A=(\mu_a,e_p)$. It
was used by Bost and Connes \cite{Bost95},\cite{Cohen99} to build
a remarkable quantum statistical model undergoing a phase
transition.

\subsection{The \lq\lq quantum Fechner law" and its thermodynamics}

Let us now make use of the concepts of quantum statistical
mechanics. Given an observable Hermitian operator $M$ and a
Hamiltonian $H_0$ one has the evolution $\sigma_t(M)$ versus time
$t$
\begin{equation}
\sigma_t(M)=e^{itH_0}Me^{-itH_0}, \label{evolution}
\end{equation}
and the expectation value of $M$ at inverse temperature
$\beta=\frac{1}{kT}$ is the unique Gibbs state
\begin{equation}
$\rm{Gibbs}$(M)=$\rm{Trace}$(M e^{-\beta
H_0})/$\rm{Trace}$(e^{-\beta H_0}). \label{Gibbs}
\end{equation}
For the more general case of an algebra of observables A, the
Gibbs state is replaced by the so-called Kubo-Martin-Schwinger (or
$\rm{KMS}_{\beta}$) state. One introduces a $1$-parameter group
$\sigma_t$ of automorphisms of $A$ so that for any $t \in R$ and
$x \in A$
\begin{equation}
\sigma_t(x)=e^{itH_0}x e^{-itH_0}, \label{evolutionA}
\end{equation}
but the equilibrium state remains unique only if some conditions
regarding the evolution of the operators $x\in A$ are satisfied:
the so-called $\rm{KMS}$ conditions \cite{Bost95}. It happens
quite often that there is a unique equilibrium state above a
critical temperature $T_c$ and a coexistence of many equilibrium
states at low temperature $T<T_c$. There is thus a spontaneous
symmetry breaking at the critical temperature $T_c$. A simple
example is the phase diagram for a ferromagnet. At temperature
larger than $T\simeq 10^3 K$ the disorder dominates and the
thermal equilibrium state is unique, while for $T<T_c$ the
individual magnets tend to align each other, which in the three
dimensional space $\mathcal{R}^3$ of the ferromagnet yields a set
of extremal equilibrium phases parametrized by the symmetry group
$SO(3)$ of rotations in $\mathcal{R}^3$.

In Bost and Connes approach the Hamiltonian operator is given as
the logarithm of the number operator $H_0=\ln N$, i.e. it is a
kind of \lq\lq quantum Fechner law" since the \lq\lq sensor"
operator $H_0$ is the logarithm of the \lq\lq physical" operator
$N$, in contrast to the ordinary Hamiltonian $H_0=N+\frac{1}{2}$
of the harmonic oscillator. The dynamical system $(A,\sigma_t)$ is
defined by its action on the Fock states
\begin{equation}
H_0|n\rangle= \ln n |n\rangle. \label{Hamilton}
\end{equation}
Using the relations $e^{-\beta H_0}|n\rangle=e^{-\beta\ln
n}|n\rangle=n^{-\beta}|n\rangle$, it follows that the partition
function of the model at the inverse temperature $\beta$ is
\begin{equation}
$\rm{Trace}$(e^{-\beta H_0})=\sum_{n=1}^{\infty}
n^{-\beta}=\zeta(\beta),~\Re{\beta}>1. \label{zeta}
\end{equation}
where $\zeta(\beta)$ is the Riemann zeta function introduced at
the end of Sect.\ref{classical}. Applying (\ref{evolutionA}) to
the Hamiltonian $H_0$ in (\ref{Hamilton}) one gets
\begin{equation}
\sigma_t(\mu_a)=a^{it}\mu_a~~$\rm{and}$~~\sigma_t(e_p)=e_p.
\label{evolutionOp}
\end{equation}
One observes that under the action of $\sigma_t$ the additive
phase operator (\ref{phaseop2}) is invariant, while the
multiplicative shift operator (\ref{shift2}) oscillates in time at
angular frequency $\ln a$. It is found in
\cite{Bost95},\cite{Cohen99} that the pole $\beta=1$ of the
Riemann zeta function separates two dynamical regimes.

In the high temperature regime $0<\beta \le1$ there is a unique
$\rm{KMS}_{\beta}$ state of the dynamical system $(A,\sigma_t)$.
The expectation value $\psi_{\beta}(\frac{p}{q})$ for the
irreducible fractions $\frac{p}{q}$, $(p,q)=1$ is
\begin{equation}
\psi_{\beta}(\frac{p}{q})=\prod_{\stackrel{b~\rm{prime}}{k_b \neq
0 }}b^{-k_b\beta}\frac{1-b^{\beta -1}}{1-b^{-1}},
 \label{KMShighT}
\end{equation}
and the $k_b$ are the exponents in the unique prime decomposition
of the denominator $q$
\begin{equation}
q= \prod_{p~ \rm{prime}}p^{k_p}.\label{primedec}
\end{equation}

In the low temperature regime $\beta>1$ things become more
involved. The $\rm{KMS}_{\beta}$ state is no longer unique, as it
is in the case of the low temperature ferromagnet. The symmetry
group $G$ of the dynamical system ($A,\sigma_t$) is a subgroup of
the automorphisms of $A$ which commutes with $\sigma_t$
\begin{equation}
w\circ \sigma_t=\sigma_t\circ w,~\forall w\in G,~\forall t\in
\mathcal{R},
\end{equation}
where $\circ$ means the composition law in $G$. When the
temperature is high the system is disordered enough that there is
no interaction between its constituents and the equilibrium state
of the system does not see the action of the symmetry group $G$,
which explains why the thermal state is unique. At lower
temperature $T<T_c$ the constituents of the system may interact.
The symmetry group $G$ then permutes a family of extremal
$\rm{KMS}_{\beta}$ states generating the possible states of the
system after phase transition. The symmetry group for the algebra
$A=(\mu_a,e_p)$ and the \lq\lq quantum Fechner Hamiltonian" $H_0$
in (\ref{Hamilton}) is the Galois group
$G=Gal(\mathcal{Q}^{\rm{cycl}}/\mathcal{Q})$ of the cyclotomic
extension $\mathcal{Q}^{\rm{cycl}}$ of $\mathcal{Q}$. The field
$\mathcal{Q}^{\rm{cycl}}$ is the one generated on $\mathcal{Q}$ by
all the roots of unity. The Galois group $G$ is the automorphism
group of $\mathcal{Q}^{\rm{cycl}}$ which preserves the sub-field
$\mathcal{Q}$. It is isomorphic to
$\mathcal{Z}_q=(\mathcal{Z}/q\mathcal{Z})^*$ which is the group of
integers modulo $q$ with the zero element removed.

One immediately sees that the action of the Galois group $G$
commutes with the $1$-parameter group defined in
(\ref{evolutionOp}). Hence $G$ permutes the extremal
$\rm{KMS}_{\beta}$ states of $(A,\sigma_t)$. To describe these
thermal states for $\beta>1$ one starts from the action of
operators $\mu_a$ and $e_p$ on Fock states as given in
(\ref{shift2}) and (\ref{phaseop2}), and one allows the
permutation by the symmetry group $G$. For each $w\in G$, one has
a representation $\pi_w$ of the algebra $A$ so that
\begin{equation}
\pi_w(\mu_a)|n\rangle =|an($\rm{mod}$~q)\rangle, \nonumber \\
\end{equation}
\begin{equation}
\pi_w(e_p)|n\rangle=w(\exp(2i\pi n\frac{p}{q})|n\rangle,
\end{equation}
and it is shown in \cite{Bost95} that the state
\begin{equation}
\phi_{\beta,w}(x)=\zeta(\beta)^{-1}$\rm{Trace}$(\pi_w(x)e^{-\beta
H_0}), ~x\in A
\end{equation}
is a $\rm{KMS}_{\beta}$ state for $(A,\sigma_t)$. The action of
$G$ on $A$ induces an action on these thermal states which
permutes them.
% and the map is homeomorphism (a continuous $1$ to
%$1$ invertible relation) of $G$ onto a set of extremal thermal
%states.
 For $\beta>1$ the expectation value acting on the phase
operator $e_p$
may be written as %
$\rm{KMS}(e_p)=q^{-\beta}\prod_{\stackrel{p~ \rm{divides}~ q }{p~
\rm{prime}}}\frac{1-p^{\beta-1}}{1-p^{-1}}$. But it can be also
found that formula (\ref{KMShighT}) is valid in the whole range of
temperatures $0 \le \beta \le \infty$.

Plotting the thermal state (Fig. 3) one observes the following
asymptotes. In the high temperature limit $\beta$=0 the thermal
state tends to $1$; in the low temperature range $\beta \gg 1$ the
thermal state is well approximated by the function
$\mu(q)/\phi(q)$, where the M\"{o}bius function $\mu(q)$ is $0$ if
the prime decomposition of $q$ contains a square, $1$ if $q=1$ and
$(-1)^k$ if $q$ is the product of $k$ distinct primes. Close to
the critical temperature at $\beta=1$ the fluctuations are
squeezed and the thermal state is quite close to
$\epsilon\Lambda(q)/q$, when $\epsilon=1-\beta \rightarrow 0$,
where $\Lambda(q)$ is the Mangoldt function defined in
(\ref{Mangoldt}). It is an unexpected surprise that phase
fluctuations at the critical region of the quantum perception
model resembles the ones of the classical phenomenological model
of time measurements.

It is shown in our earlier papers \cite{Planat03}, in eq. (38)
that the normalized M\"{o}bius function $\mu(q)/\phi(q)$ is the
dual to the Mangoldt function $\Lambda(n)$ when one projects on to
the basis of Ramanujan sums. Ramanujan sums $c_q(n)$, which are
defined as the sums of $n^{\rm{th}}$ powers of primitive roots of
unity, generalize the cosine function, being quasi-periodic in $n$
and aperiodic in $q$. The thermal states in the transition region
can thus be considered as an unfolding through the dimensions $q$
of the unique high temperature thermal state. This also suggests
to consider the inverse temperature $\beta$ and the dimension $q$
of the Hilbert space as complementary to each other in the
Heisenberg sense.

\subsection{Time perception and memory recording}

We tried to make clear that time perception is the quantum
counterpart to time measurements. The quantum algebra of shift and
clock operators together with the \lq\lq quantum Fechner law" lead
to well defined thermal perception states, that should play some
role in memory encoding and conscious activity. There seems to be
a growing consensus that the mental activity in brain doesn't
relate to classical synchronization rules in neurons, but to some
other non local mechanism at a smaller scale, possibly at the
mesoscopic scale of microtubules within neurons
\cite{Jibu94},\cite{Nanopoulos95},\cite{Rosu97}.

Quantum field theory of many body systems and the mechanism of
spontaneous breakdown of symmetry was already proposed to
accommodate some features of the conscious brain function. Memory
recording was represented by the ordering induced in the ground
state by the condensation of modes after symmetry breaking of the
rotational symmetry of the electrical dipoles of the water
molecules. The recall mechanism was described by the excitation of
these collective modes from the ground state under the action of
an external input similar to the one which was previously produced
by memory encoding. In a dissipative extension of the model two
modes squeezed coherent states of the radiation field were found
to play a role in extending considerably the memory capacity
\cite{Vitiello95}.

The set of collective modes in the ground sate of Vitiello's model
is replaced in our model by the set of thermal phase states (the
extremal $\rm{KMS}_{\beta}$ states). Transitions among the vacua
are replaced by transitions among the $\rm{KMS}_{\beta}$ states
under the action of the cyclotomic group. Vitiello's model can be
seen as an attempt to include the environment in the quantum model
of brain, thus the idea of two entangled modes, one in the brain,
the other one in the environment. Our approach seems more general
being able to describe the environment thanks to a continuous
temperature parameter $\beta$, and putting on the scene more
general squeezed thermal states at the transition temperature. The
discrete parameter $q$ is the dimension of the Hilbert space of
Fock states. It plays the role of a bandwidth in conscious and
unconscious activity. A very wide bandwidth $q \rightarrow \infty$
would mean meditation, a small bandwidth would be associated to
precise memory encoding, jumps between dimensions, with tiny
energy needed, thanks to the ridges of the $\rm{KMS}$ versus $q$
and $\beta$ diagram (see. Fig. 3). Another closely related view of
time perception based on cyclic properties of Galois Fields
$GF(b^k)$ has been proposed \cite{Saniga98}. This pencil model
sheds new light on profoundly distorted perceptions of time
characterizing a number of mental psychoses, drug-induced states,
as well as many other \lq\lq altered" states of consciousness. The
high temperature thermal state could be associated to the ordinary
sense of time, and the low temperature thermal states to altered
states of time perception.

\begin{figure}[ht]
\centering{\resizebox{18cm}{!} {\includegraphics{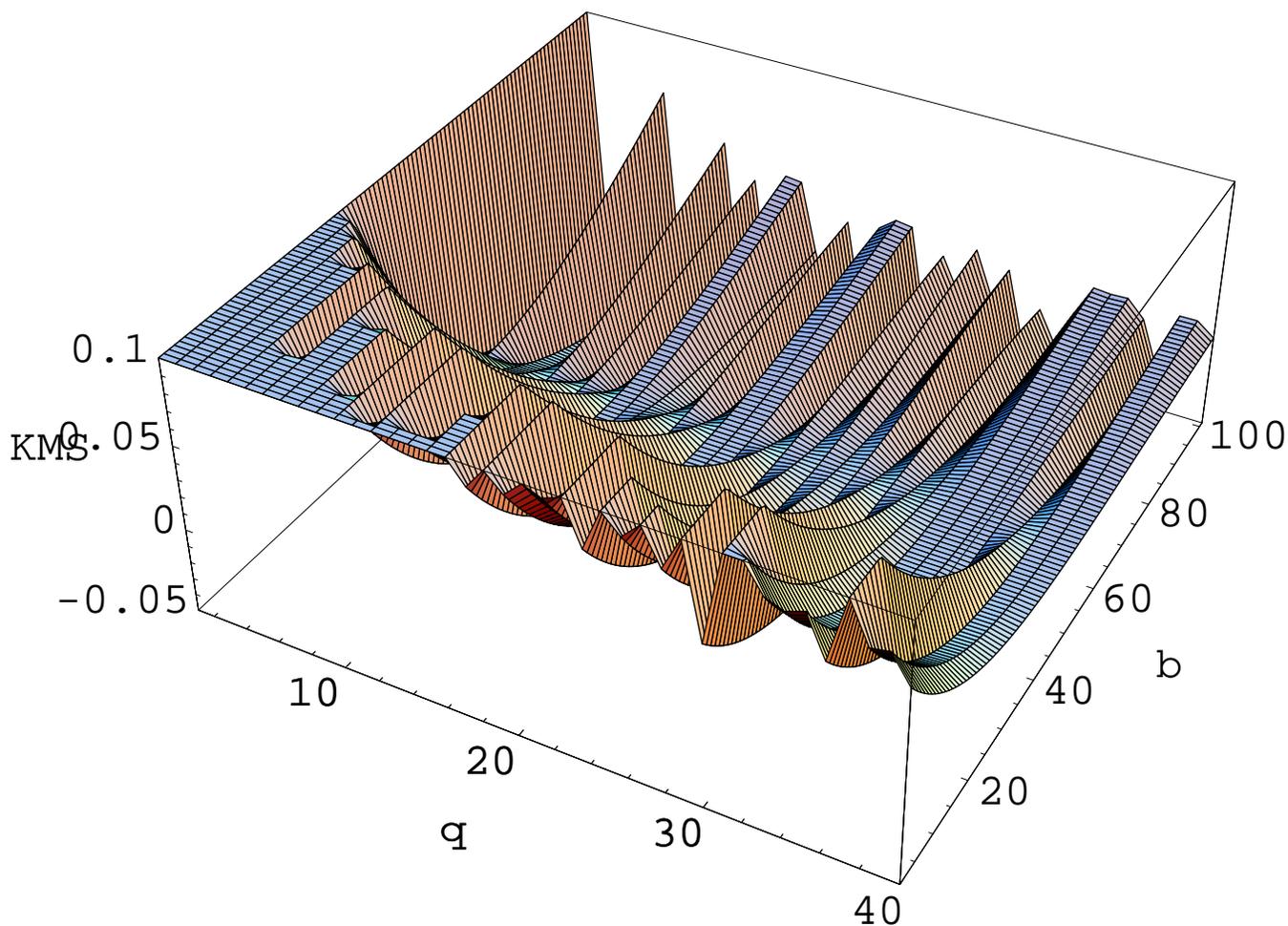}}}
\caption{The KMS thermal state as given from (\ref{KMShighT}). The
inverse temperature scale is from $\beta=0.5$ to $1.5$. The
dimension is from $q=1$ to $40$  as shown.  Some parts of the
graphics are truncated in the high temperature region $\beta<1$.
Thermal fluctuations are clearly reduced at the inverse critical
temperature $\beta=1$. }
\end{figure}

\section*{Acknowledgments}
The author wish to thank Serge Perrine, Metod Saniga, Haret Rosu
and Naoum Daher for their critical reading of the manuscript.


\section*{References}
\begin{thebibliography}{99}

\bibitem{Flanagan03} B. Flanagan. Are perception fields quantum
fields? NeuroQuantology 2003; 3:334-364.

\bibitem{Planat03} M. Planat. Invitation to the \lq\lq spooky"
quantum phase-locking effect and its link to $1/f$ fluctuations.
Preprint 2003; arXiv:quant-ph/0310082.

\bibitem{Schroeder99} M.~R. Schroeder. Number Theory in Science and
Communication. Berlin: Springer 1999; 16-24.

\bibitem{Gilden01} D.~L. Gilden. Cognitive emissions of $1/f$
noise. Psychological Review 2001; 108:33-56.

\bibitem{Poincare05} H. Poincar\'{e}. Mathematical magnitude and experiment.
Chapter 2 in Science and Hypothesis. London: Walter Scott
Publishing 1905; 17-34.

\bibitem{Bost95} J. B. Bost and A. Connes. Hecke algebra, type III factors and phase transitions with
spontaneous symmetry breaking in number theory.  Selecta
Mathematica 1995;  1: 411-457.

%\bibitem{Planat01}  M. Planat. $1/f$ noise, the measurement of
%time and number theory. Fluc. and Noise Lett. 2001; 1: R65.


\bibitem{Lynch95} R. Lynch. The quantum phase: a critical review. Physics Reports 1995;
 256: 367-436.

\bibitem{Vourdas93} A. Vourdas. Phase states: An analytic approach
in the unit disk. Physica Scripta 1993; T48: 84-86.

\bibitem{Perelomov86} A. Perelomov. Generalized Coherent States
and Their Applications. Berlin: Springer Verlag 1986.

\bibitem{Bouchiat98} V. Bouchiat, D. Vion, P. Joyez, D. Est\'{e}ve and
M. H. Devoret.  Quantum coherence with a single Cooper pair.
Physica Scripta 1998;  1: 165-170.

 \bibitem{Moree03}  P. Moree P and H. Hommerson H.  Value distribution of Ramanujan sums
and of cyclotomic polynomial coefficients. Preprint 2003;
math.NT/0307352

\bibitem{Ekert00} A. Ekert, P. Hayden and H. Inamori. Basic
concepts in quantum computation. in Coherent Atomic Matter Waves.
Les Houches Summer School. Berlin: Springer 2001. Preprint 2000;
quant-ph/0011013.

\bibitem{Cohen99} P.~B. Cohen. Dedekind zeta functions and quantum
statistical mechanics. Contemporary mathematics 1999; 241:
121-128.

\bibitem{Jibu94} M. Jibu, S. Hagan, S.~R. Hameroff, K. H. Pribam
and K. Yasue. Quantum optical coherence in cytoskeletal
microtubules: implications for brain function. Biosystems 1994;
32: 195-209.

\bibitem{Nanopoulos95} D.~V. Nanopoulos. Theory of brain function,
quantum mechanics and superstrings. CERN-TH 1995; 95-128. Preprint
1995; hep-ph/9505374.

\bibitem{Rosu97} H. Rosu. Essay on mesoscopic and quantum brain. Metaphysical Review 1997;
3:1-12. Preprint 1997; gr-qc/9409007.

\bibitem{Vitiello95} G. Vitiello. Dissipation and memory capacity
in the quantum brain model. Int. J. Mod. Phys. 1995; 9: 973-989.

\bibitem{Saniga98} M. Saniga, Pencils of conics: a means towards a deeper understanding of the arrow of time?
Chaos, Solitons and Fractals 1998; 9: 1071-1086. Also: Temporal
dimension over Galois fields of characteristic two. Chaos,
Solitons and Fractals 1998; 9: 1095-1104.


\end{thebibliography}
\end{document}